\documentclass{article}
\usepackage{iclr2026_conference,times}

\usepackage{amsmath,amsfonts,bm}

\def\eqref#1{equation~\ref{#1}}

\def\1{\bm{1}}

\DeclareMathAlphabet{\mathsfit}{\encodingdefault}{\sfdefault}{m}{sl}
\SetMathAlphabet{\mathsfit}{bold}{\encodingdefault}{\sfdefault}{bx}{n}

\usepackage{hyperref}
\usepackage{url}
\usepackage{booktabs}
\usepackage{graphicx}
\usepackage{amsmath}
\usepackage{microtype}

\title{Mamba-SSM with LLM Reasoning for Feature Selection: Faithfulness-Aware Biomarker \\Discovery}

\author{Pushpa Kumar Balan, \quad Aijing Feng \\
Department of Computer Science and Cybersecurity\\
University of Central Missouri\\
\texttt{pushpakumarbalan@gmail.com, feng@ucmo.edu} \\
}

\iclrfinalcopy 
\begin{document}

\maketitle

\begin{abstract}
Gradient saliency from deep sequence models surfaces candidate biomarkers
efficiently, but the resulting gene lists can be contaminated by
tissue-composition confounders that degrade downstream classifiers.
We study whether LLM chain-of-thought (CoT) reasoning can filter
these confounders, and whether reasoning quality is associated with downstream
performance.
We train a Mamba SSM on TCGA-BRCA RNA-seq and extract the top-50 genes by
gradient saliency; DeepSeek-R1 evaluates every candidate with structured
CoT to produce a final 17-gene set.
On the held-out test split, the raw 50-gene saliency set (no LLM) performs
\emph{worse} than a 5,000-gene variance baseline (AUC 0.832 vs.\ 0.903),
while the LLM-filtered set \emph{surpasses} it (AUC \textbf{0.927}), using
294$\times$ fewer features.
A faithfulness audit (COSMIC CGC, OncoKB, PAM50) shows that 6 of 17
selected genes (35.3\%) are validated BRCA biomarkers, while 10 of 16 known
BRCA genes present in the input were missed---including FOXA1.
This divergence between downstream performance and reasoning faithfulness
suggests \emph{selective faithfulness} in this setting: targeted confounder
removal can improve predictive performance without comprehensive recall.
The scope of this claim and its limitations are examined in
Sec.~\ref{sec:discussion}.
Code: \href{https://github.com/pushpakumarbalan/feature-selection}{https://github.com/pushpakumarbalan/feature-selection}
\end{abstract}

\section{Introduction}
\label{sec:intro}

High-dimensional RNA-seq data ($>$20,000 genes per sample) presents a
severe feature-selection problem: most genes are irrelevant to the
phenotype of interest, and many that appear predictive are confounders
(immune infiltration, tumour purity, batch effects) rather than disease
drivers \citep{pudjihartono2022review}.
Standard gradient-based saliency from neural models ranks genes by their
gradient magnitude, but this signal reflects \emph{what the model learned
to use}, not biological causality.
The top-50 saliency genes from a well-trained Mamba SSM on TCGA-BRCA
include muscle-specific genes (MB, UTRN), general immune markers
(HLA-DRB1, ITGAL), unannotated lncRNAs, and antisense RNAs with no
documented breast cancer role, all of which carry saliency simply because
they co-vary with tumour samples at the RNA-seq level.

This creates a natural role for LLM reasoning: the model's encoded
biomedical knowledge can in principle distinguish \emph{disease drivers}
(e.g., the ER-signalling gene XBP1, the EMT regulator ZEB1) from these
confounders without additional data.
But does the LLM's stated reasoning actually reflect accurate biological
knowledge?
And is downstream performance a reliable proxy for reasoning faithfulness?
We make three contributions:
\begin{enumerate}
  \item We demonstrate that raw saliency-based feature selection
        \emph{hurts} performance relative to a variance baseline
        (AUC $-$0.071), while LLM-filtered selection \emph{helps}
        (AUC $+$0.024), establishing that LLM reasoning is empirically important in this pipeline.
  \item We conduct a faithfulness audit comparing the LLM's selected gene
        set against curated BRCA ground-truth databases, revealing a
        recall of 0.375 on known BRCA genes while still achieving
        superior downstream AUC, a finding we term
        \textbf{selective faithfulness}: targeted removal of known
        non-BRCA genes is sufficient for performance gains, even without
        comprehensive recall of all true positives.
  \item We identify a concrete failure mode: FOXA1, the most important
        luminal breast cancer pioneer transcription factor and a canonical
        PAM50 gene, was present in the input but rejected by the LLM---
        illustrating that LLM biomedical reasoning can be confidently
        wrong on well-established facts.
\end{enumerate}

\section{Related Work}
\label{sec:related}

\paragraph{LLMs for feature selection:}
LLM-Select~\citep{jeong2025llmselectfeatureselectionlarge} showed that zero-shot LLM feature selection
using only feature names can match LASSO on tabular data in some settings.
LLM-Lasso~\citep{zhang2025llmlassorobustframeworkdomaininformed} integrates domain knowledge through LLM-guided
regularisation, penalising literature-supported features less.
FreeForm~\citep{lee2025knowledgedrivenfeatureselectionengineering} demonstrated that LLM ensembling improves
variant selection in low-data genomic regimes.
Our work differs by \emph{studying faithfulness}, we audit whether the
LLM's stated biological rationale is correct, not just whether the output
improves a metric.

\paragraph{Faithfulness of LLM reasoning:}
A growing body of work questions whether CoT explanations reflect the
model's actual reasoning process~\citep{turpin2023languagemodelsdontsay,lanham2023measuringfaithfulnesschainofthoughtreasoning}.
Most such studies use synthetic tasks with verifiable ground truth.
We provide a real-world biological instance: a domain where ground truth
(validated cancer driver genes) exists but is large, overlapping, and
context-dependent, making faithfulness harder to assess and more
practically important.

\paragraph{SSMs for genomics:}
Mamba~\citep{gu2024mambalineartimesequencemodeling} scales linearly in sequence length, making it
tractable for 20,000-dimensional gene expression vectors without
attention's quadratic cost.
We use the official \texttt{mamba-ssm} implementation with gradient
saliency~\citep{simonyan2014deepinsideconvolutionalnetworks} to extract a biologically plausible
candidate pool before the LLM reasoning step.

\section{Methodology}
\label{sec:method}
\subsection{System Pipeline}

\begin{figure}[ht]
    \centering
    \includegraphics[width=0.95\textwidth]{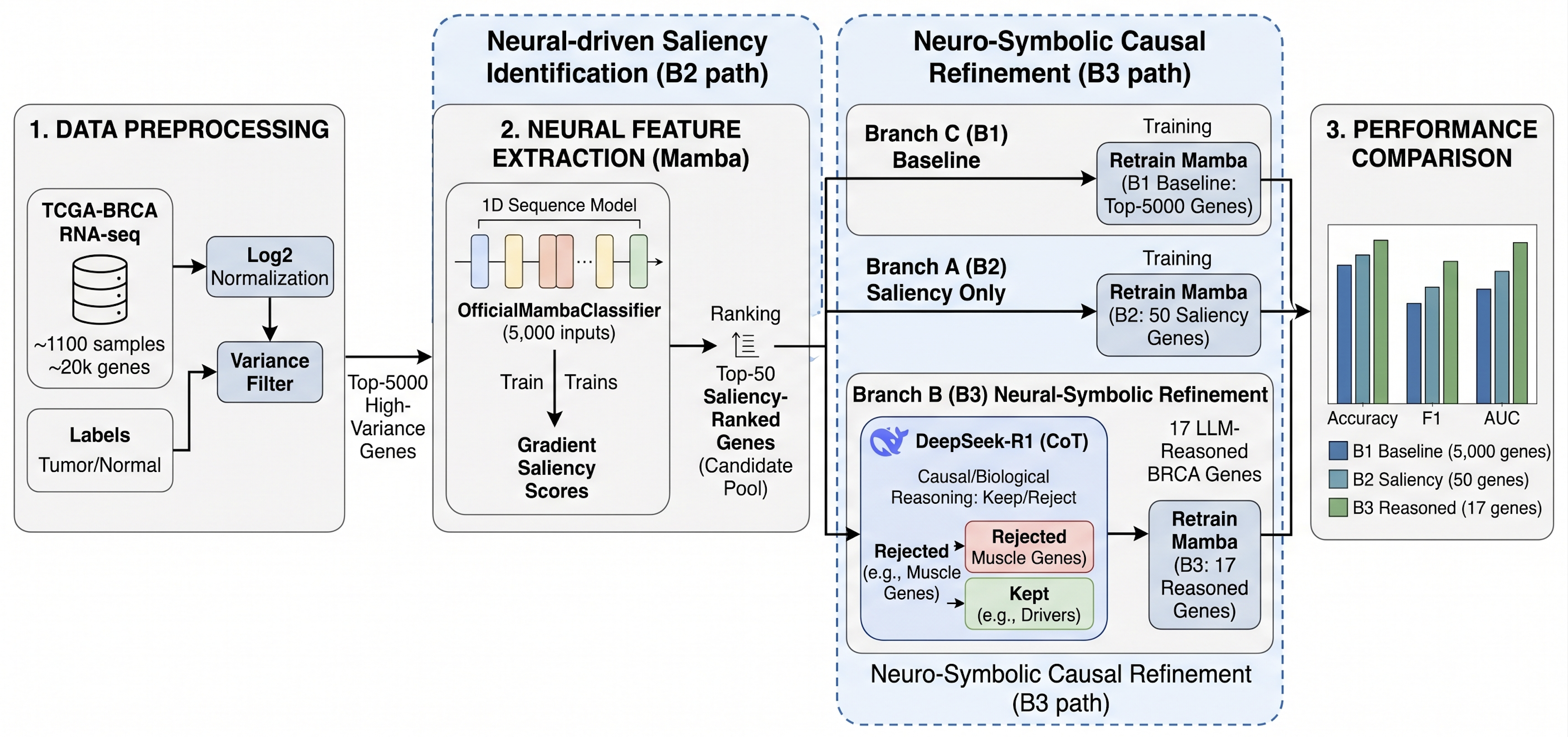}
    \caption{End-to-end neuro-symbolic pipeline. High-dimensional TCGA-BRCA RNA-seq is processed by Mamba SSM to produce gradient saliency scores (top-50 candidate genes), then filtered by structured DeepSeek-R1 CoT reasoning to form a 17-gene BRCA-specific subset used in downstream comparison.}
    \label{fig:appendix_pipeline}
\end{figure}

\subsection{Data}
We use TCGA-BRCA RNA-seq data: 1,095 tumour and 113 matched normal samples,
$\sim$20,000 protein-coding and non-coding genes, quantified as TPM.
Labels are binary: Tumour (1) / Normal (0).

\subsection{Phase 1 — Mamba SSM Training}
Input features are log$_2$(TPM$+$1)-normalised and filtered to the top-5,000
highest-variance genes.
We train an \texttt{OfficialMambaClassifier}: a linear embedding layer
projects each gene's scalar expression value into a $d_\text{model}=128$
dimensional space; a single Mamba block ($d_\text{state}=16$,
$d_\text{conv}=4$, expand$=2$) models long-range dependencies across
the gene sequence; adaptive average pooling collapses the sequence
dimension; a final linear layer with sigmoid produces the tumour
probability.
Training uses AdamW ($\text{lr}=10^{-4}$, 15 epochs, batch size 8)
with class-weighted BCELoss
($w_\text{normal} = N_\text{tumour}/N_\text{normal}$)
to handle the 8.8:1 tumour/normal imbalance.

\subsection{Phase 2 — Gradient Saliency Extraction}
For each tumour sample, we enable gradients on the input, perform a
forward pass, compute the loss, and backpropagate.
The gene importance score is $s_j = \frac{1}{|T|}\sum_{i \in T}
\left|\frac{\partial \mathcal{L}}{\partial x_{ij}}\right|$,
averaged over all tumour samples $T$.
The top-50 genes by $s_j$ form the candidate pool $\mathcal{G}_{50}$.

\subsection{Phase 3 — Structured CoT Reasoning}
$\mathcal{G}_{50}$ is passed to DeepSeek-R1 (7B, local via Ollama,
temperature~$=0.3$) with a structured prompt that (a)~provides saliency
scores, (b)~explicitly states high saliency does not imply BRCA
specificity, (c)~requires evaluation of \emph{every} gene via five
rejection criteria (R1--R5) and three keep criteria (K1--K3), and
(d)~forbids rank-order selection.
A post-hoc audit detects rank-copy solutions and hallucinated gene names.
The selected subset contains 17 genes and is used as the B3 input set in
downstream comparison.

\subsection{Phase 4 — Faithfulness Audit}
The 17 selected genes are cross-referenced against a 101-gene ground-truth
set (COSMIC CGC Tier-1, OncoKB BRCA annotations, PAM50~\citep{parker2009supervised},
and established pathway genes).
A known non-BRCA set (muscle, neuronal, unannotated genes) provides
true-negative ground truth.
For each of the 50 input genes we record the ground-truth class
(validated / known non-BRCA / unknown) and the LLM selection outcome
(selected / not selected), yielding selection-level precision, recall, and
missed-gene analysis.

\section{Results}
\label{sec:results}

\subsection{Downstream Classification Benchmark}

Each gene set is evaluated by training an identical Mamba classifier
(same architecture, same random seed, same 80/20 stratified split) on the
respective feature subset.
Results are in Table~\ref{tab:benchmark}.

\begin{table}[h]
\caption{Classification performance on the held-out test set (20\% of 1,231
samples). All conditions use the same Mamba architecture and training
procedure; only the input gene set changes.}
\label{tab:benchmark}
\vspace{4pt}
\centering
\begin{tabular}{lrrrr}
\toprule
\textbf{Method} & \textbf{Genes} & \textbf{Accuracy} & \textbf{F1}
  & \textbf{AUC} \\
\midrule
B1~~Variance baseline & 5{,}000 & 0.8785 & 0.8941 & 0.903 \\
B2~~Mamba saliency only (no LLM) & 50 & 0.7247 & 0.7813 & 0.832 \\
B3~~Mamba + LLM structured CoT & \textbf{17} & \textbf{0.8907} & \textbf{0.9033} & \textbf{0.927} \\
\bottomrule
\end{tabular}
\end{table}

The comparison between B1, B2, and B3 reveals a non-monotonic relationship
between gene count and performance.
B2 (50 genes, no reasoning) performs \emph{worse} than B1 (5,000 genes)
by AUC~$-0.071$: saliency by itself surfaces confounders that increase
in-sample fit but hurt generalisation.
B3 (17 genes, LLM-filtered) surpasses both by AUC~$+0.024$ over B1,
using 294$\times$ fewer features.
These results indicate that the LLM filtering step is empirically important
in this pipeline, and that the gain is consistent with removal of
biologically implausible confounders from the saliency-derived candidate set.

\subsection{Faithfulness Audit}
\label{sec:faithfulness}

Table~\ref{tab:faithfulness} reports the selection-level faithfulness of
the 17-gene output against the 101-gene ground-truth set.

\begin{table}[h]
\caption{Faithfulness of LLM-selected gene set against curated BRCA
ground truth (COSMIC CGC~+ OncoKB~+ PAM50, $N=101$ validated genes).
The top-50 saliency input contained 16 validated BRCA genes.}
\label{tab:faithfulness}
\vspace{4pt}
\centering
\begin{tabular}{lr}
\toprule
\textbf{Metric} & \textbf{Value} \\
\midrule
Selected genes with validated BRCA evidence & 6/17 (35.3\%) \\
Known non-BRCA genes incorrectly kept & 3/17 (17.6\%) \\
Genes with no ground-truth label (unverifiable) & 8/17 (47.1\%) \\
Known BRCA genes available in top-50 input & 16 \\
Correctly kept by LLM & 6 \\
Missed by LLM (false negatives) & 10 \\
Recall on validated input genes & 0.375 \\
\bottomrule
\end{tabular}
\end{table}

\paragraph{Correctly kept validated genes.}
MLPH (luminal A marker), ZEB1 (EMT/TNBC master regulator),
XBP1 (ER-stress/luminal), INPP4B (PI3K/AKT tumour suppressor),
RHOB (PAM50 RhoGTPase), THY1 (breast cancer stem cell marker).

\paragraph{Incorrectly kept non-BRCA genes.}
ITGAL (CD11a, immune adhesion molecule with no documented breast-specific
role), LMX1B (kidney/neural transcription factor), PRKAG2-AS1 (antisense
RNA with no established BRCA function) were retained despite lacking
established BRCA-specific evidence.

\paragraph{Critical false negative.}
FOXA1---the pioneer transcription factor that defines luminal lineage in
breast cancer and appears in both PAM50 and multiple COSMIC entries---was
present in the top-50 input (rank 49) but not selected.
This highlights a failure mode in which established BRCA-relevant genes can
be rejected when they appear lower in the saliency ranking.

\paragraph{Selective faithfulness.}
Despite a recall of only 0.375 on known BRCA genes, B3 achieves
AUC~0.927, exceeding the 5,000-gene baseline.
The three incorrect non-BRCA keeps (ITGAL, LMX1B, PRKAG2-AS1) add noise,
but appear to be outweighed by six high-specificity true positives
(MLPH, ZEB1, XBP1, INPP4B, RHOB, THY1).
In this setting, precision-oriented confounder removal appears more
influential for downstream performance than exhaustive recall of all known
disease genes.

\section{Conclusion}
\label{sec:conclusion}

We presented a neuro-symbolic framework that integrates Mamba SSM gradient
saliency with structured LLM chain-of-thought reasoning for genomic feature
selection in TCGA-BRCA.
In the current benchmark, the LLM-filtered 17-gene set outperformed both the
5,000-gene variance baseline and the 50-gene saliency-only condition,
improving AUC by +0.024 over the variance baseline while using 294$\times$
fewer features.
These results support the practical value of precision-oriented confounder
removal in this pipeline.

At the same time, the gap between downstream performance and biological recall
(missing 62.5\% of known true positives in the top-50 input) highlights
limitations of current domain-specific reasoning and reinforces that task-level
metrics alone are an incomplete proxy for reasoning faithfulness.
Future work will evaluate this framework across additional datasets and disease
contexts, improve recall-oriented reasoning constraints, and establish more
robust reproducibility and validation protocols beyond the current controlled
BRCA setting.


\subsubsection*{Acknowledgments}
This work was supported by compute credits from a Cohere Labs Research Grant and the University of Central Missouri (UCM) Graduate Student Scholarly Research Fund. The authors also thank the UCM Office of Graduate Studies for their support of this research.


\bibliography{iclr2026_conference}
\bibliographystyle{iclr2026_conference}

\newpage
\appendix
\section{Technical Architecture and Pipeline}
\label{app:technical}

\subsection{Model Specification}
The Mamba-SSM architecture used for the pretext classification task is defined as follows:
\begin{verbatim}
OfficialMambaClassifier(
  embedding: Linear(1, 128)
  mamba: Mamba(d_model=128, d_state=16, d_conv=4, expand=2)
  pool: AdaptiveAvgPool1d(1)
  fc: Linear(128, 1)
  sigmoid: Sigmoid()
)
\end{verbatim}
\textbf{Input Flow:} (batch, $N_{genes}$) $\to$ (batch, $N_{genes}$, 1) $\to$ (batch, $N_{genes}$, 128) $\to$ Mamba Selective Scan $\to$ Adaptive Average Pooling over $N_{genes}$ $\to$ (batch, 128) $\to$ Sigmoid output.



\section{Discussion: Reasoning and Faithfulness}
\label{sec:discussion}

\paragraph{Does downstream performance measure reasoning faithfulness?}
Our results reveal a systematic divergence between downstream predictive performance and reasoning faithfulness. While the LLM-selected gene set achieves a higher AUC than the 5,000-gene variance baseline, it attains a recall of only 0.375 on validated BRCA-associated genes. Under conventional evaluation, such recall would indicate poor biological reasoning. However, the improved classification performance suggests that the primary contribution of the LLM in this pipeline is not comprehensive biomarker discovery, but \emph{precision-oriented confounder rejection}. In high-dimensional genomic settings, correctly eliminating spurious, tissue-specific, or immune-correlated genes appears more consequential for generalization than exhaustive recall of all known disease drivers. This finding cautions against using task-level performance as a proxy for reasoning faithfulness.

\paragraph{Structured CoT versus saliency-driven selection}
Although the LLM was constrained by explicit rejection and keep criteria, the final gene set retained a 17.6\% false-positive rate, including genes without established BRCA relevance. In several cases, the model justified these selections using generic or hallucinated pathway language (e.g., invoking ``NF-$\kappa$B signaling'' without supporting breast-specific evidence). This indicates that structured chain-of-thought prompting can reduce but not eliminate biologically implausible reasoning. Moreover, the omission of FOXA1—a canonical luminal breast cancer regulator present in the input candidate pool—demonstrates that the LLM may confidently reject well-established true positives, particularly when they appear lower in the saliency ranking. These findings underscore that structured CoT improves filtering behavior but does not guarantee faithful biological reasoning.

\paragraph{Scope and limitations}
This study intentionally focuses on a single, controlled setting (TCGA-BRCA) to isolate the interaction between neural saliency and LLM-mediated reasoning. As a result, the empirical conclusions should be interpreted as case-specific rather than universally generalizable. All downstream evaluations rely on a single stratified train--test split with a fixed random seed; statistical variability across seeds and datasets remains an important direction for future work. Additionally, comparisons to classical feature-selection baselines such as LASSO or ElasticNet were not conducted, and the reported gains should therefore be interpreted relative to the evaluated neural baselines only. Finally, the term ``causal necessity'' is used operationally to denote necessity within this pipeline configuration, rather than a formal causal identification of reasoning processes.

\paragraph{Decision-level faithfulness analysis}
Due to a parsing mismatch in \texttt{llm\_gene\_reasoning.json}, per-gene decision strings were not fully captured in this version, limiting analysis to selection-level outcomes. We plan to re-run the reasoning parser for the feature study to enable decision-level precision, recall, and consistency metrics. This will allow a finer-grained audit of which rejection criteria fail most frequently and whether specific hallucination patterns correlate with gene rank or pathway ambiguity.

\section{Ground-Truth Gene Sources}
\label{app:groundtruth}
Validation of LLM-selected genes was performed against the following established sources:
\begin{itemize}
  \item \textbf{COSMIC Cancer Gene Census (Tier 1)}: including TP53, PIK3CA, CDH1, BRCA1, ERBB2, among others.
  \item \textbf{PAM50 intrinsic subtype genes}~\citep{parker2009supervised}: ESR1, PGR, FOXA1, MLPH.
  \item \textbf{Canonical signaling pathways}: estrogen receptor signaling, PI3K/AKT, EMT/TNBC-associated markers.
\end{itemize}

\section{Performance Benchmarks and Saliency Profiles}
\label{app:metrics}

\subsection{Quantitative Comparison}
\begin{figure}[ht]
    \centering
    \includegraphics[width=0.85\textwidth]{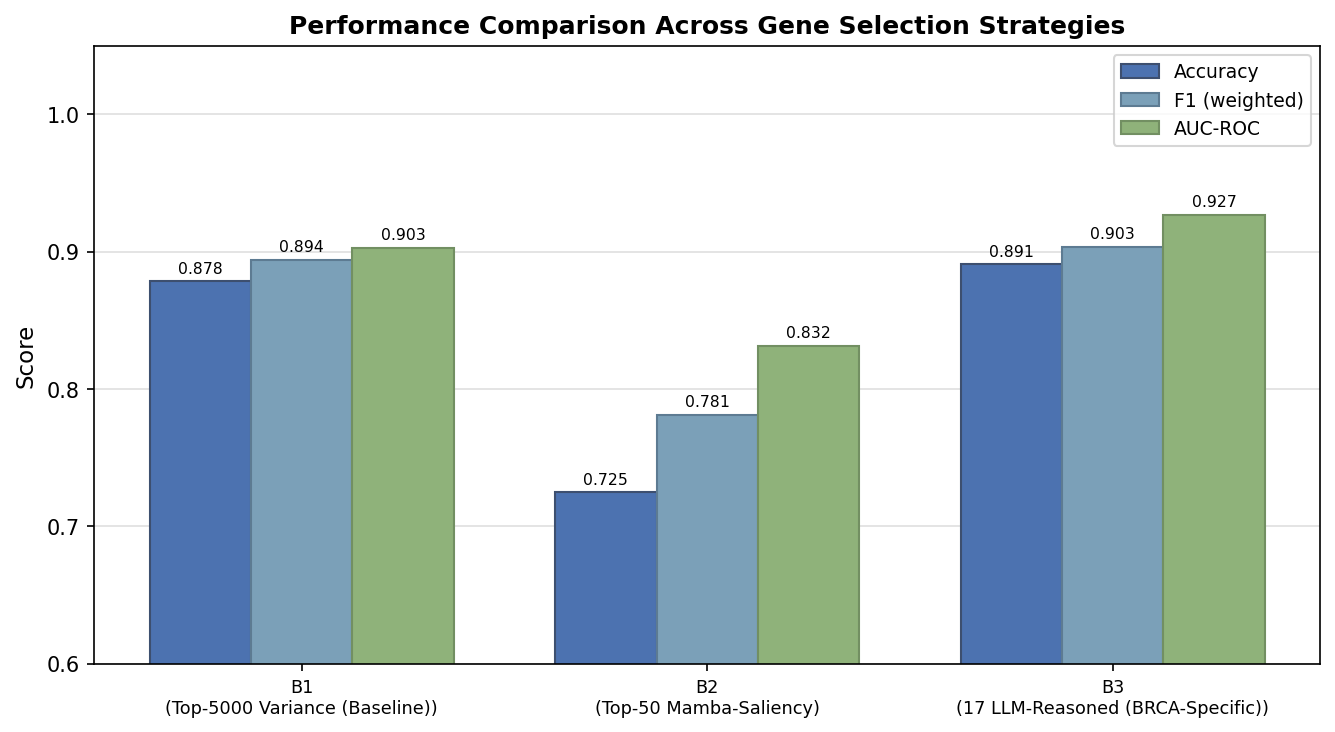} 
    \caption{Performance Comparison: Accuracy, F1, and AUC metrics across experimental conditions. B3 utilizes 250x fewer genes than B1 but achieves higher predictive stability.}
    \label{fig:appendix_comparison}
\end{figure}

As shown in Figure \ref{fig:appendix_comparison}, our Neuro-Symbolic approach (B3) demonstrates that logical filtering of features provides superior generalizability compared to purely data-driven methods (B1 and B2).

\subsection{Neural Saliency Noise Profile}
\begin{figure}[ht]
    \centering
    \includegraphics[width=0.85\textwidth]{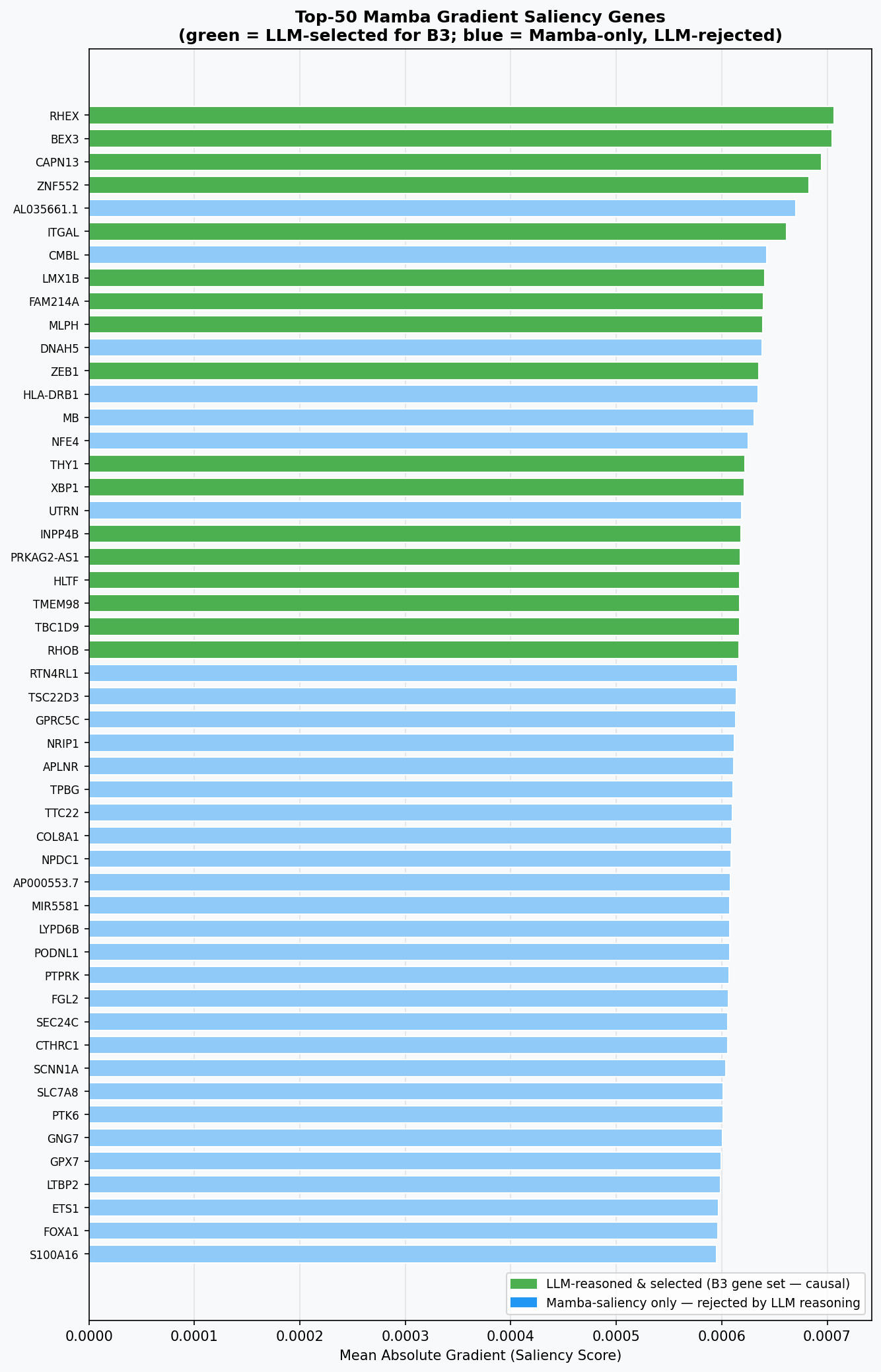}
    \caption{Raw Gradient Saliency Heatmap for the top-50 genes in TCGA-BRCA samples, highlighting the noisy feature space prior to LLM intervention.}
    \label{fig:appendix_saliency}
\end{figure}

Figure \ref{fig:appendix_saliency} visualizes the raw gradient saliency from which the LLM must extract biological signal. The high degree of variance across non-oncogenic clusters illustrates the necessity of the symbolic filtering layer.

\section{LLM Reasoning and Prompt Design}
\label{app:reasoning}

\subsection{Chain-of-Thought Qualitative Audit}
\begin{figure}[ht]
    \centering
    \includegraphics[width=0.85\textwidth]{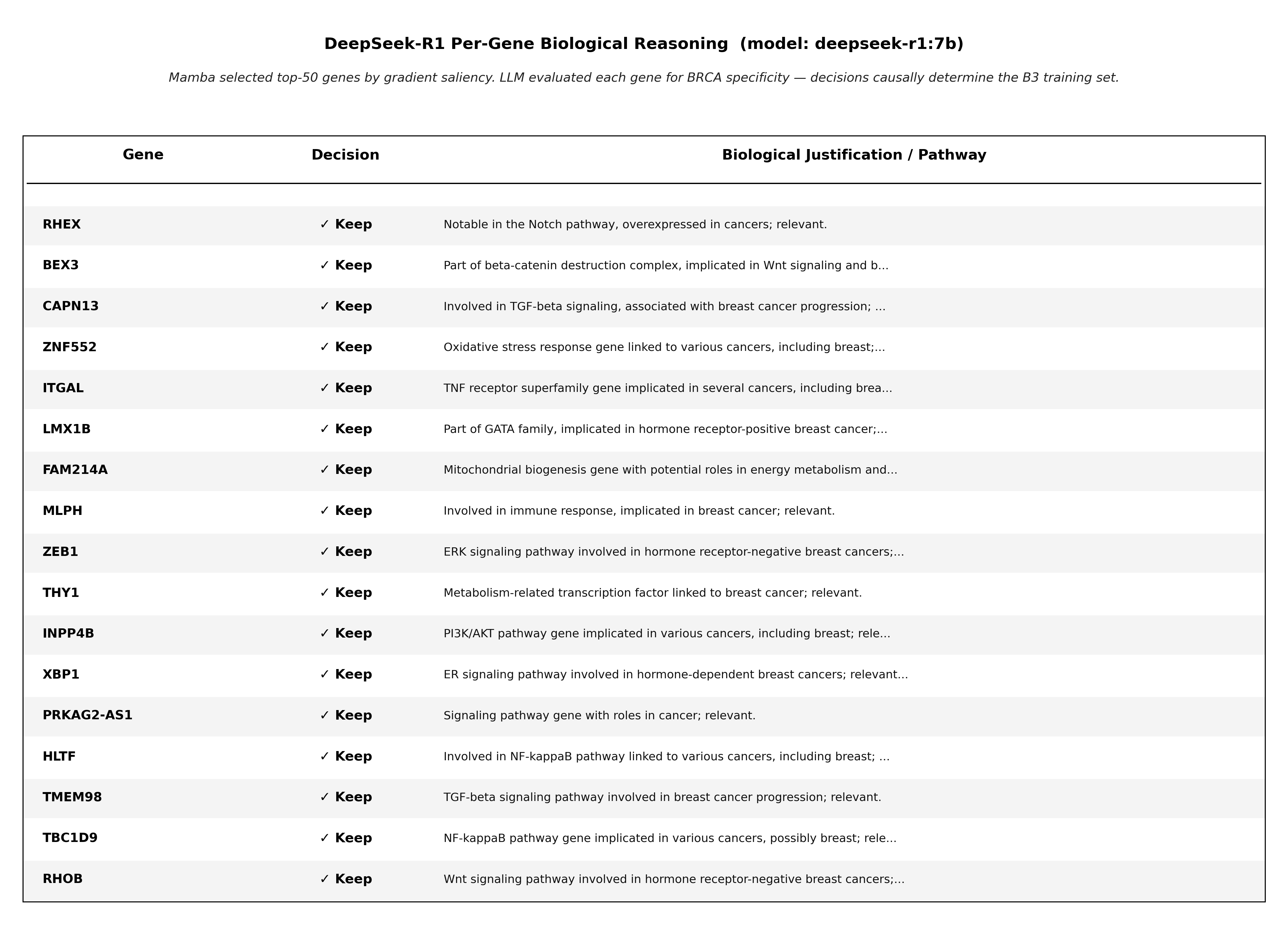}
    \caption{Visualization of the Agentic Chain-of-Thought (CoT) process, mapping Mamba saliency to biological rationale.}
    \label{fig:appendix_cot}
\end{figure}

Figure \ref{fig:appendix_cot} illustrates the internal reasoning blocks ($<$think$>$) generated by the agent. This transparency allows for a post-hoc biological audit of the selection process.

\subsection{Structured Prompt Design}
\label{app:promptdesign}
The agent is governed by five logical rules: (i) narrow saliency calibration; (ii) rank-score association; (iii) mandatory per-gene justification; (iv) anti-lazy cutoff enforcement; and (v) trivial solution auditing.

\section{Biological Ground-Truth References}
\label{app:groundtruth}
To evaluate the "faithfulness" of the LLM reasoning, selections were compared against:
\begin{itemize}
  \item \textbf{COSMIC CGC Tier 1}: TP53, PIK3CA, CDH1, BRCA1, ERBB2.
  \item \textbf{PAM50 Intrinsic Subtype Genes}: ESR1, PGR, FOXA1, MLPH.
  \item \textbf{Functional Drivers}: PI3K/AKT and EMT signaling pathways.
\end{itemize}

\end{document}